\documentclass[12pt]{article}

\usepackage{setspace}
\usepackage{amsmath}
\usepackage{graphicx}

\usepackage{amsfonts}
\usepackage{amssymb}
\usepackage[german,english]{babel}
\usepackage{bm,pstricks}                          

\usepackage{setspace}
\usepackage{calligra}

\begin{document}

\title{Frequency thermal response and cooling performance in a microscopic system with a time-dependent perturbation.}

\maketitle

\author{ Natalia Beraha$^{1,2,*}$, Alejandro Soba $^{1,2}$,
 M. Florencia Carusela$^{1,2}$\\
\textnormal{$^{1}$ Instituto de Ciencias, Universidad Nacional de Gral. Sarmiento, Los Polvorines, Argentina\\
$^{2}$ Consejo Nacional de Investigaciones Cient\'ificas y T\'ecnicas, Buenos Aires, Argentina\\
$^{*}$Email: nberaha@ungs.edu.ar}}


\begin{abstract} 

Following the nonequilibrium Green's function formalism we study the thermal transport in a composite chain subject to a time-dependent perturbation. The system is formed by two finite linear asymmetric harmonic chains subject to an on-site potential connected together by a time-modulated coupling. The ends of the chains are coupled to two phononic reservoirs at different temperatures. We present the relevant equations used to calculate the heat current along each segment. We find that the system presents different transport regimes according the driving frequency and temperature gradients. One of the regimes corresponds to a heat pump against thermal gradient, thus a characterization of the cooling performance of the device is presented.

\end{abstract}

\section{Introduction}

 Nowadays, the technological implementations in the meso and nanoscale require the management of substantial energies that can be generated, becoming harmful for a device. So significant improvements are needed in the direction of controlling energy, and thus heat, to avoid structural damages. However, many studies show that the management of heat can also display intriguing features which allow the design of devices with novel operating regimes \cite{RenLi,ren2}.

Spontaneously heat flows from objects at high temperatures to objects at lower temperatures. However heat pumps enables heat to flow against temperature gradient by means of an applied external work. 
At the meso and nanoscale several interesting applications have been developed in molecular electronics, thermometry and thermal machinary
\cite{nitz,wang,pek}.  
 Many models of heat pumps have been proposed based on different mechanisms such as heat ratchets that periodically adjusts two baths'  temperatures while the average remains equal,  brownian heat motors to shuttle heat across the system \cite{LZHL},  heat pumps which directs heat against thermal bias in nanomechanical systems \cite{AMCC}. 
At molecular levels phonon pumps can also be induced by an external force or by mechanical switch on-off of the coupling between different parts of the system \cite{ nuestroPA,  HY, AWL, zhang, CW, Tratch,Extratch}.  Experimentally the last can be done in molecular junctions or in molecular systems, for example, varying the distance among them or applying stretchings and compressions.

Cooling is another relevant feature because of applications in the quantum realm.   Experimental implementations of quantum refrigerators have been developed based on electronic devices in the presence of ac driving fields \cite{cp1, cp2, cp3, cp4}, absortion of phonons (heat) with electrons using nanomechanical devices or active feedback for cooling nanomechanical cantilevered beams \cite{feed1, feed2, feed3, feed4}.  
In turn models based on pumping phenomena have been proposed as moving barriers in a cavity to pump phonons from a cold reservoir to a hotter one, or driven two level systems or molecular junctions in contact with phononic baths \cite{cool1, AR}.
In the present work, the mechanism underlying cooling is mainly the phonon manipulation that can work both in insulators and electrical conductors, unlike the electronic cooling where charge, spin and coherence are important.

We present a microscopic model of a phonon pump based on a time dependent modulation of the contact between two one dimensional components.  We show that this system can work not only as a heat engine, but it can also operate as a phononic refrigerator.

The paper is organized as follows. In section $2$ we present the microscopic model. In section $3$ we describe the theoretical framework and methodology used to solve numerically the problem. In section $4$ we present the different regimes for heat transport, discussing the role of external frequency and the size of system. In section $5$ we analyze the cooling regime comparing with the optimal case. The last section is devoted to summarize the results.

\section{The model}

\begin{figure}[h]
\centering
\includegraphics[scale=0.08]{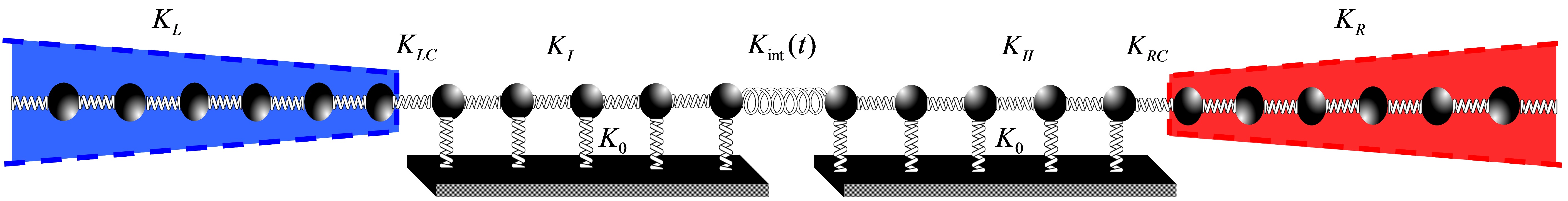} 
\caption{\label{fig:Figura1} Sketch of the microscopic model. The system is composed by two one-dimensional chains of particles coupled by a modulated interaction in time. The semi-infinite left chain is kept at a temperature $T_L$ and the semi-infinite right one at $T_R$.}
\end{figure}

We propose a microscopic model for energy transport assisted by acoustic phonons. We consider a one dimensional array of atoms, harmonically and bidirectionally coupled, sketched in Fig.\ref{fig:Figura1}.  
The chain is divided in two segments $I$ (left) and $II$ (right) with different coupling intensities $K_{I}$ and $K_{II}$ between identical atoms or molecules (referred as "masses") and coupled together also harmonically with a coupling constant $K_{int}(t)$. The system is subject to a local harmonic pinning potential and connected to $L$ (left end) and $R$ (right end) semi-infinite chains of masses $m_L$ and $m_R$ respectively, coupled harmonically by spring constants $K_L$ and $K_R$ . This chains play the role of phonon reservoirs with temperatures $T_L$ and $T_R$.
We assume that the system can only vibrate longitudinally that is, we are modelling a heat pump assisted only by longitudinal acoustic modes.

The Hamiltonian of the system can be written as: $H=H_{L}+H_{R} + H_{central}+H_{int}(t)+H_{contact}$
where $H_{L, R}$ are the Hamiltonians of the left (L)/right (R) reservoirs respectively,  $H_{central}$ describes the two central chains (I and II), $H_{int}$ represents the interaction between them and $H_{contact}$ represents the contact between the two central chains and the reservoirs.

\begin{equation}
H_{central}= \sum_{\alpha = I, II} \sum_{i=1}^{N_\alpha-1} \frac{p_{i,\alpha}^2}{2m_{i,\alpha}}+\frac{1}{2} K_{\alpha}(x_{i+1,\alpha}-x_{i,\alpha})^2 +\frac{1}{2} K_{0} x_{i,\alpha}^2
\end{equation}
with $N_\alpha$ the total number of  particles. 

The two segments $\alpha =I , II$ of the central chain have equal length $N_\alpha=N/2$. $K_{\alpha}$ are the elastics constants in each segment and $K_{0}$ is the spring constant of the local pinning potential.  They are harmonically coupled with a time dependent strength ${K}_{int}(t) = K_{int}^{0}+K_{int}^{1}(t)=K_{0}(1+ cos(\omega_{0}t))$.  Thus the interaction Hamiltonian can be written as:
\begin{equation}
H_{int}= \frac{1}{2} K_{int}(t)(x_{N_{1},I}-x_{N_{1},II})^2
\end{equation}

with $m_{i, \alpha}$ the mass of the $ith$ atom in the chain $\alpha$, $x_{i,\alpha}= q_{i,\alpha}-i a$ denotes the displacement from the equilibrium position $i a$, where $a$ is the equilibrium distance between particles and $p_{i, \alpha}$ is the momentum.  The index $i=1$ labels the atom of a segement that is in contact with a reservoir

The reservoir Hamiltonians corresponding to the semi-infinity chains are
\begin{equation}
H_{\beta}= \sum_{i=1}^{N_\beta} \frac{p_{i,\beta}^2}{2m_{i,\beta}}+\frac{1}{2} K_{\beta}(x_{i,\beta}-x_{i+1,\beta})^2 
\end{equation}
with  $N_\beta \rightarrow \infty$ ($\beta = L,R$) number of  particles.

It is convenient to express this Hamiltonian in terms of normal modes for open boundary conditions
\begin{equation}
H_{\beta}= \sum_{n=0}^{N_\beta} \frac{p_{n,\beta}^2}{2m_{n,\beta}}+\frac{1}{2} K_{\beta}[1- cos(u_{n,\beta}) ]x_{n,\beta}^2 
\end{equation}

The corresponding transformations are:
\begin{equation}
x_{i,\beta}= \frac{2}{N_{\beta}+1} \sum_{n=0}^{N_\beta} sin(u_{n,\beta} i) x_{n,\beta}
\end{equation}

and

\begin{equation}
p_{i,\beta}= \frac{2}{N_{\beta}+1} \sum_{n=0}^{N_\beta} sin(u_{n,\beta} i) p_{n,\beta}
\end{equation}

with

\begin{equation}
u_{n,\beta}= \frac{n \pi}{N_\beta +1},    n=0,......,N_\beta
\end{equation}
 
The contact is described by the Hamiltonian

\begin{equation}
H_{contact}=  \frac{1}{2} K_{Lc}(x_{1, L}-x_{1, I})^2+ \frac{1}{2} K_{Rc}(x_{1,R}-x_{1, II})^2 
\end{equation}

where elements $(1, I)$ and $(1, II)$ are the sites of the central chain in contact with the L/R bath.


In order to study the energy current that flows between the system and the reservoirs, we focus on the energy balance.  Conservation of energy implies that the total power invested by all the external fields or mechanical agents must be dissipated into the reservoirs at a rate given by:

\begin{equation}
\bar{P}_{total}=\bar{J}_{L}+ \bar{J}_{R}
\label{power}
\end{equation}

where $\bar{J}_{L}$ and $\bar{J}_{R}$ are the rates of heat absorbed by reservoir $L$ and $R$ respectively (defined positive when heat flows into the reservoir).   In the case of periodic fields or drivings  "$\bar{...}$" means time average over one period.

The energy current that flows to and out a given elementary volume should cancel each other when the system reaches the steady state and in the case there is not local power injection, since there are not temporal variations of the mean local density of energy.  As our Hamiltonians involve only nearest-neighbors terms, the minimum volume that we consider is one that encloses nearest-neighbors sites. From the continuity equation and energy conservation, the local time-dependent heat current for the incoming energy current from connecting site $1,\beta$ of the central chain towards each reservoir $\beta$ gives:

\begin{equation}
J_{\beta}(t)= \sum_{n}^{N_\beta} k_{n,\beta}\left\langle x_{\beta,n} \dot x_{1,\beta}\right\rangle  
\label{cur1}
\end{equation}
with $k_{n\beta}=K_{\beta}\sqrt{\frac{2}{\frac{N_{\beta}}{2}+1}}sin(u_{n}^\beta)$.

We define the net heat current in each segment $\bar{J} _{\beta}$ averaging over an integer number of periods after a transient time as:

\begin{equation}
\bar{J}_{\beta}=\frac{1}{\tau} \int_{0}^{\tau} J_{\beta}(t) dt 
\end{equation}

In the steady state the value of the current is independent of the site $i$ in each segment $\alpha$ of the central part, therefore $J_{i, I} = J_{L}$ and $J_{i, II} = J_{R}$.

%

\section{Theoretical framework: Non-equilibrium Green's function method and Dyson equation.}

Following the procedure define in \cite{Arrach}, we define the greater, lesser, advanced and retarded Green's functions

\begin{equation}
G^{>}_{k, k'}(t,t') = \textit{i}  \left\langle x_{k}(t) x_{k' }(t')\right\rangle
\end{equation}

\begin{equation}
G^{<}_{k, k'}(t,t') = \textit{i}  \left\langle x_{k'}(t') x_{k}(t)\right\rangle
\end{equation}

\begin{align}
G^{R}_{k, k'}(t,t') &=  -\textit{i} \Theta(t - t') \left\langle [x_{k}(t) x_{k'}(t')]\right\rangle  \\ 
                               &= \Theta(t - t')  [G^{<}_{k, k'}(t,t') - G^{>}_{k, k'}(t,t')]
\end{align}

\begin{align}
G^{A}_{k, k'}(t,t') &=  -\textit{i} \Theta(t - t') \left\langle [x_{k}(t) x_{k'}(t')]\right\rangle  \\ 
                               &= \Theta(t' - t)  [G^{<}_{k, k'}(t,t') - G^{>}_{k, k'}(t,t')]
\end{align}

The Dyson equation for the retarded function corresponding to the central chains (without the contacts) is:

\begin{multline}
-\left[ \partial^{2}_{t'}+F_{k,k'}(t) \right] G^{R}_{k, k'} (t,t')+F_{k,k'+1}(t) G^{R}_{k, k'+1} (t,t') F_{k,k'-1}(t) G^{R}_{k, k'-1} (t,t')=\\
=\frac{1}{n}\delta_{k,k'} \delta (t-t')
\end{multline}

In turn, when writing the Dyson equation for the retarded Green's function along the contacts, it is possible to integrate the degrees of freedom of the reservoirs, then:

\begin{equation}
- \left[\partial^{2}_{t'} G^{R} (t,t')+ G^{R}(t,t') \emph{ F(t')} \right] - \int ds \Sigma^{R}(t,s)  G^{R}(s,t') = \frac{1}{m}\delta_{k,k'} \delta (t-t')
\label{DysonEq}
\end{equation}

The force matrix includes the interparticle interactions, the pinning interaction and the system-reservoir interaction and is defined as : $\emph{F}(t) = \emph{F}^{0}(t)+\emph{Fo}^{0}(t)+\emph{F}^{1}(t)$, where the matrix elements are:

\begin{equation}
\emph{F}^{0}_{k,k'}(t) = \left\{ \begin{array}{cc} 
\frac{K_{I}}{m}(2 \delta_{k,k'}-\delta_{k',k\pm1}) &  1<k<N/2 \\  \\
\frac{K_{II}}{m}(2 \delta_{k,k'}-\delta_{k',k\pm1}) &  N/2+1<k<N \\ \\
\frac{K_{I}+K_{LC}}{m}\delta_{k,k'}- \frac{K_{I}}{m} \delta_{k',k\pm1}&  k=1 \\ \\
\frac{K_{II}+K_{RC}}{m}\delta_{k,k'}- \frac{K_{II}}{m} \delta_{k',k\pm1} &  k=N \\ \\
\frac{K_{I}+K_{int}^{0}}{m}\delta_{k,k'}-\frac{K_{I}+K_{int}^{0}}{m} \delta_{k',k\pm1} &  k=N/2 \\ \\
\frac{K_{II}+K_{int}^{0}}{m}\delta_{k,k'}-\frac{K_{II}+K_{int}^{0}}{m} \delta_{k',k\pm1} &  k=N/2+1
\end{array}\right. 
\end{equation}

\begin{equation}
\emph{Fo}^{0}_{k,k'}(t) = - K_{0} \delta_{k,k'} 
\end{equation}


\begin{equation}
\emph{F}^{1}_{k,k'}(t) = \left\{ \begin{array}{cc} 
\frac{K_{I}+K_{int}^{1}}{2}\delta_{k,k'}-\frac{K_{I}+K_{int}^{1}}{2} \delta_{k',k\pm1} &  k=N/2 \\ \\
\frac{K_{II}+K_{int}^{1}}{2}\delta_{k,k'}-\frac{K_{II}+K_{int}^{1}}{2} \delta_{k',k\pm1} &  k=N/2+1
\end{array}\right. 
\end{equation}

On the other hand, in Eq.(\ref{DysonEq}) $\Sigma^{R}_{k,k'}(t,t')$ corresponds to the the self energy:

\begin{equation}
\Sigma^{R}_{k,k'}(t,t')=\sum_{\beta=L,R}\delta_{k,k'}\delta_{k',k_{\beta}}\int_{-\infty}^{\infty}\frac{d\omega}{2\pi}
e^{-i \omega (t-t')} \times 
\int_{-\infty}^{\infty}\frac{d\omega}{2\pi}\frac{\Gamma_{\beta}(\omega')}{\omega- \omega' + i \eta}
\end{equation}

with $\eta$ a positive infinitesimal and $  \Gamma_{\beta}(\omega)$ the spectral density of the reservoir  $\beta$ :

\begin{multline}
  \Gamma_{\beta}(\omega)=\lim_{N_{\beta \rightarrow \infty}}\frac{2\pi\left(\frac{K_{\beta}}{m_{\beta}}\right)^2}{N_{\beta}+1} 
  \sum_{n=0}^{N_{\beta}} sin^2\left(u_{n}^{\beta}\right)\frac{1}{E_{\beta,n}} 
  \left[ \delta\left( \omega-E_{\beta,n}\right) +\delta\left( \omega+E_{\beta,n}\right) \right] =\\
  sgn\left(\omega\right)\left(\frac{K_{\beta}}{u_{\beta}}\right)^2 \Theta\left[1-\left(\frac{K_{\beta}-u_{\beta}\omega^2}{K_{\beta}}\right)^2\right]
  \sqrt{1-\left(\frac{K_{\beta}-\omega_{\beta}\omega^2}{K_{\beta}}\right)^2}
\end{multline}

where $E_{\beta, n}=\sqrt{\frac{K_{\beta}}{u_{\beta}}\left(1-cos\left(u_{n}^{\beta}\right)\right)}$.

When Langreth's rules are applied recursively, we can obtain the Dyson's equation for the lesser Green's function corresponding to coordinates along the central chain:

\begin{equation}
 G_{k,k'}^<\left(t,t'\right)=\sum_{\beta}\int{dS_{1}}\int{dS_{2}} G_{k,1\beta}^R\left(t,S_{1}\right)\textstyle\sum_{\beta}^<\left(S_{1}-S_{2}\right) G_{1\beta,k'}^A\left(S_{2},t'\right)
\label{LGFK}
\end{equation}

and to one coordinate on the chain and the other on the reservoir.

\begin{multline}
  G_{k,u_{n}^\beta}^<\left(t,t'\right)=-\frac{K_{\beta}}{u_{\beta}}\int{dS_{1}}
  [ G_{k,1\beta}^R\left(t,S_{1}\right)g_{u_{n}^\beta}^<\left(S_{1}-t'\right)+\\
  + G_{k,1\beta}^<\left(t,S_{1}\right) g_{u_{n}^\beta}^A\left(S_{1}-t'\right)]
  \label{LGFu}
\end{multline}

The corresponding lesser Green's function of the uncoupled reservoirs is:

\begin{equation}
g_{u_{n}^\beta}^<(\omega)=\frac{i\pi n_\beta (\omega)}{E_{\beta,n}} \left[\delta(\omega-E_{\beta,n})+\delta(\omega+E_{\beta,n})\right]
\end{equation}
with a Fourier transform:
\begin{equation}
\Sigma^{<}_{\beta} (\omega)=i n_{\beta}(\omega)\Gamma_{\beta}(\omega) 
\end{equation}
with $n_\beta(\omega)$ being the Bose-Eistein distribution, dependent on temperature $T_\beta $ of reservoir $\beta$.

The main goal is to obtain the exact function $D^R$, solving Eq.(\ref{DysonEq}).

Following the strategy described  in \cite{AR}, we first perform the Fourier transform with respect the "delayed time" (t') in $G^R(t,t')$.
\begin{equation}
 G^R(t,\omega)=\int_{-\infty}^{\infty} dt' e^{i\left(\omega+i0^+\right)\left(t-t'\right)} G^R(t,t')
\end{equation}
Then, considering the Fourier expansion of the time dependent Hamiltonian, we substitute in the Dyson's equation (Eq. (\ref{DysonEq})):
\begin{equation}
G^R(t,\omega)=G^{\left(0\right)}(\omega)+\sum^1_{\substack{k\neq 0\\-1}} e^{-i k \Omega_{o} t} G^R \left(t,\omega+k\Omega_{0} \right)
\mathcal{M}^{(1)}_{K} G^{(0)}(\omega)
\label{Greenretard}
\end{equation}
where 
\begin{equation}
G^{(0)}(\omega)=\left[\omega^2 \mathcal{I}-\mathcal{M}^{(0)}-\sum\nolimits^{R}(\omega)\right]^{-1}
\end{equation}
is the stationary component of the retarded Green's function of the central chains connected to reservoirs, when  there is no time-dependent perturbation.
Due to the temporal periodicity of $H(t)$ we can expand Eq.(\ref{Greenretard}) in terms of the Floquet components:
\begin{equation}
G^{R}(t,\omega)=\sum_k e^{-i k \Omega_{o} t} \hat{G}(k,\omega)
\label{Greenretardfloquet}
\end{equation}

Generally, to obtain an exact solution of Eq.(\ref{Greenretardfloquet}) it must be solved numerically. However under some approximations or conditions it can be solved more easily, as it is our case. As we are considering amplitudes of $H_{int}(t)$ smaller than the energies of the time independent part of $H$ and a time-dependent perturbation that contains only one harmonic $(k=\pm 1)$,  we can evaluate the Green's function up to first under as:

\begin{align}
\hat{G}(0,\omega)&=G^0(\omega)\\
\hat{G}(\pm k,\omega)&=G^0(\omega\pm k\Omega_{o})\hat{H}^{1}_{\pm k} G^{0}(\omega)
\end{align}

Now we can we calculate and express the dc component of the heat current in terms of the obtained Green's functions. Starting from Eq.(\ref{cur1}),  the time dependent heat current is:

\begin{equation}
J_{\beta}(t)=-\sum_n\frac{K_{\beta}}{m_{\beta}}\sqrt{\frac{2}{N_{\beta}+1}} sin\left(u_{n}^{\beta}\right)\lim_{t\rightarrow t'}
Re\left[i\frac{\partial}{\partial t'}G^{<}_{1{\beta}, u_{n}^{\beta}}(t,t')\right ]
\label{currentgreen} 
\end{equation}

Using the function obtained in Eq.(\ref{LGFu}) and the expansion given in Eq.(\ref{Greenretardfloquet}) we obtain the dc heat current flowing in or out of the reservoir $\beta$:

\begin{equation}
\bar{J}_{\beta}=\sum_{{\beta}'=L,R}\sum_{k=1}^{1}\int_{-\infty}^{\infty}\frac{d\omega}{2\pi} (\omega+k\Omega_{o})\left[n_{\beta'}(\omega)-n_{\beta}\omega+k\Omega_{o})\right]\Gamma_{\beta}(\omega+k\Omega_{o}) \Gamma_{\beta'}(\omega)\left|\hat{G}_{1{\beta}, 1{\beta'}}(k,\omega)\right|^2
\label{FullJ} 
\end{equation}

It is interesting to observe that the heat current may not be zero even in the absence of a temperature gradient, just mediated by the energy contribution due to the power injected into the system.

\section{Results}

In the linear chain, the energy flowing in/out of a lead can be calculated from Eq.(\ref{FullJ}). 
In a first step, we work with a modulated coupling with a strength smaller than the other coupling constants. Under this approximation the use of perturbative calculations  would be sufficiently. However as we explore broad parameter ranges where these calculations are not expected to produce accurate results, we calculate currents exactly by numerically solving the Dyson equations.


We use the following dimensionless parameters: spring constants $K_{i}$ in units of $K_{R}$, moments in units $[a(mk_{R})^{1/2})]$, frequencies in units $[(K_{R}/m)^{1/2}]$ and temperatures in $[a^2K_{R}/k_{B}]$.  For a typical atom and a typical situation these units corresponds to frequencies $\sim 10^{13}s^{-1}$ and temperatures $\sim 10^{3}K$. Thus the nondimensional temperatures $0.1$ to $1.0$ correspond to temperatures of the order $100$ to $1000K$. 

We explore several situations of operation for our device.  First, we sketch the case when strength of interaction between segments takes a constant value.  Transport is induced by the constant thermal bias between reservoirs,  that in the absence of time-dependent perturbations the dc heat current is given by the well-known Landauer-B\" uttiker formula for phononic systems \cite{AR,LB}:

\begin{equation}
J_{\alpha} = \int_{k=-\infty}^{\infty}\frac{d\omega}{2\pi} \omega \sum_{\beta=L,R} T_{\alpha \beta} [n_{\beta} (\omega)-n_{\alpha}(\omega)]
\label{Landauer}
\end{equation}

with

\begin{equation}
T_{\alpha, \beta}(\omega) =   \left| \hat{\calligra{G}}_{l_{\alpha}, l_{\beta}}^R (0,\omega)\right| ^2 \Gamma_{\alpha}(\omega)\Gamma_{\beta}(\omega)
\label{eq:Landauer}
\end{equation}

the transmission function between reservoirs $\alpha$ and $\beta$. 

\begin{figure}[h]
\centering
\includegraphics[scale=0.35]{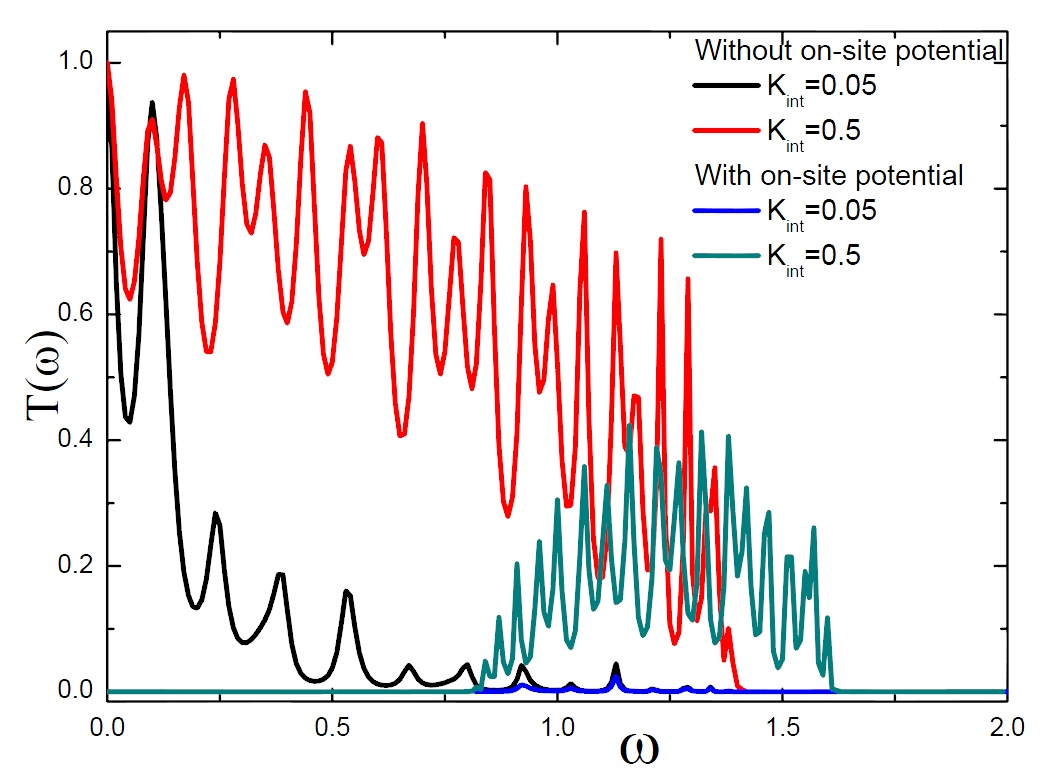} 
\caption{The transmission function $T(\omega)$ defined in Eq.(\ref{eq:Landauer}) for two segments coupled without and with an on-site potential. Red and black curves (green and blue) correspond to the case without (with) on-site potential with a strength $K_{int}=0.05, 0.5$ respectively. }
\label{fig: TLandauer}
\end{figure}

In order to show how the presence of the on-site potential and the coupling constant affects the transmission we plot in Fig.\ref{fig: TLandauer},  $T_{\alpha}(\omega)$ for a weak and strong coupling with and without pinning potential.

As our model only consider harmonic interactions, the on-site pinning potential plays a crucial role in establishing a steady-state current that can be estimated from Eq.(\ref{FullJ}).  The dc heat current corresponds to the quasi-steady state current established after a transient that is established earlier as long the on-site pinning potential is stronger \cite{CLW}.

 On the other hand the heat transmission is mainly mediated by low frequency modes, the main responsible for heat conduction.  The on-site potential produce a shift in the minimum cutoff frequency of the phonon bands from zero to $K_{pot}$,   so the transmission function takes appreciable values if $\sqrt{K_{pot}}<\omega<\sqrt{K_{pot}+4 K_{i}}$ \cite{LLW}. When the strength of the pinning potential is increased, the device has a behavior closer to a thermal insulator.  For higher values of the potential, large current oscillations can be suppressed, even for high temperatures. In Fig.\ref{fig: TLandauer} it is exemplified how the transmission function depends on the pinning potential.  The transmission for $\omega \rightarrow 0$ is independent of the value of the local coupling between segments, as it is expected for non localized modes. The system becomes a good thermal conductor. The effect of $K_{int}$ is to increase the contribution of the localized modes to the transmission along the interface. That is the contact acts as a scatter point for high frequency phonons but not for low frequency ones. These soft phonon modes (including zero-mode) play a fundamental role in the heat conduction.

In a similar way we can define from Eq.(\ref{FullJ}) the $k-th$ component of the transmission function for the perturbed system

\begin{equation}
T _k^{\pm} (\omega,\omega_0)=\Gamma_{\beta}(\omega\pm k\omega_{0}) \Gamma_{\beta'}(\omega) \left|\hat{G}_{1,{\beta}; 1,{\beta'}}(k,\omega)\right|^2
\label{shift} 
\end{equation}

\begin{figure}[h]
\centering
\includegraphics[scale=0.25]{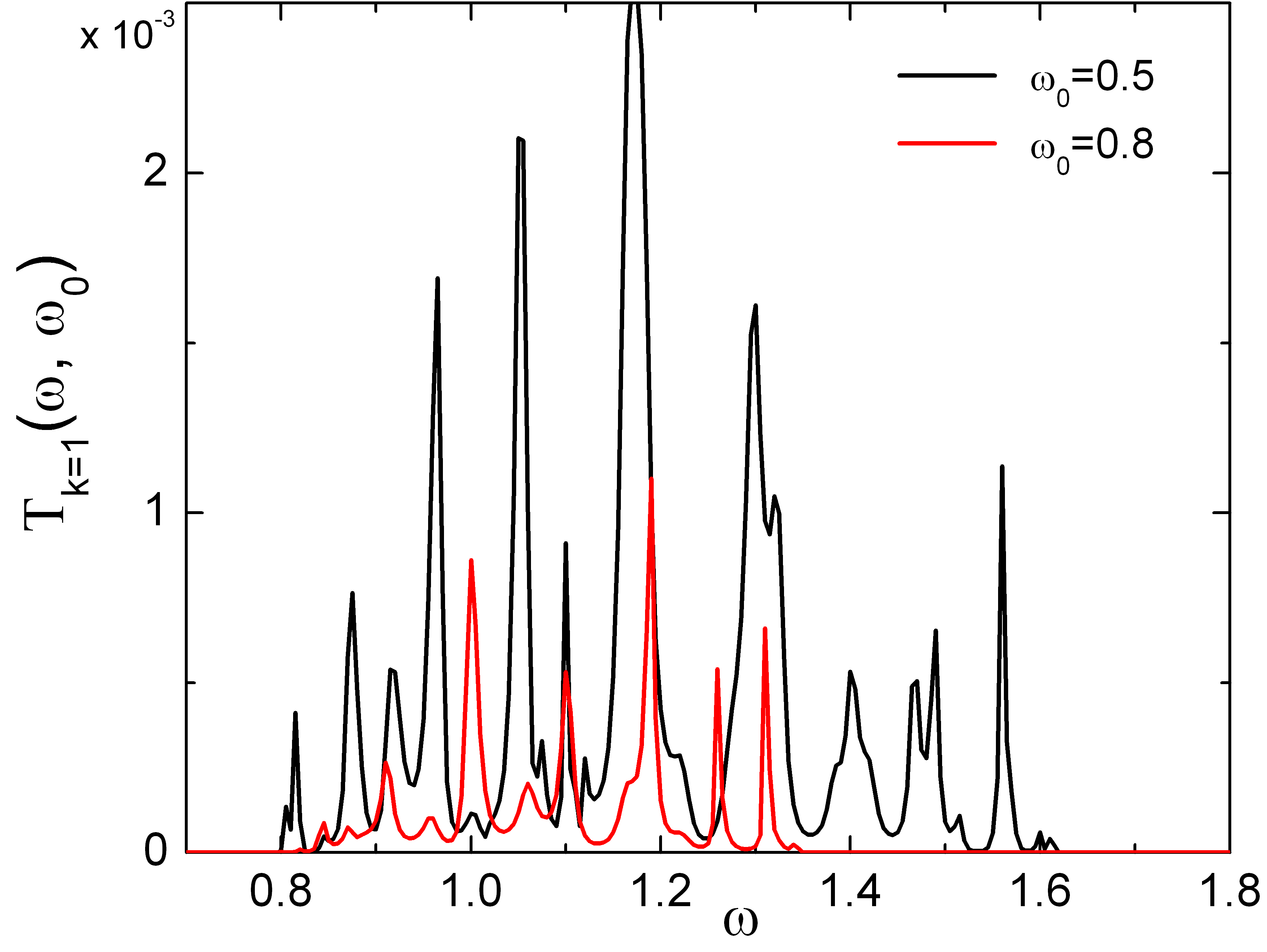}
\caption{$T^{+}_{1}(\omega,\omega_0)$ defined in Eq.(\ref{shift}) for $\omega_0=0.8, 0.5$.}
\label{fig:shift}
\end{figure}

In Fig.\ref{fig:shift} we compare the normalized $T _k^{\pm} (\omega,\Omega_0)$ for two different frequencies. Changing the frequency of modulation $\omega_0$ the phonon spectra of the $k$ channel can be narrowed or widen, thus more o less higher frequency phonons (localized) can be activated affecting the heat conduction. 

For time dependent coupling but in the limit of adiabatic driving or for very fast modulation $K_{int}(t) \propto const$ so the system reduces to a static case.
When $\omega_{0}$ is very low (adiabatic driving limit), the system reduces to two coupled segments with a coupling constant $K_{int}/2$. In the very fast oscillating limit ($\omega_{0}\rightarrow \infty$) the coupling convers to a time average constant value $K_{int}$.  Our interested is focus in the regime in between these two limits.

The transmission function shows that the system presents a multiresonant energy transport \cite{nuestroPA}. Therefore with a suitable modulation, the allowed phonon band can be changed or shifted and so the phonons involved in the transport. Consequently heat transfer along the device can be enhanced or restrict controlling and tuning {\it dynamically} the phononic thermal channels.  And in turn the contact acts as a scatter point or interface where phonons can gain or lose a phonon of frequency $\omega_0$.

\begin{figure}[h]
\centering
\includegraphics[scale=0.6]{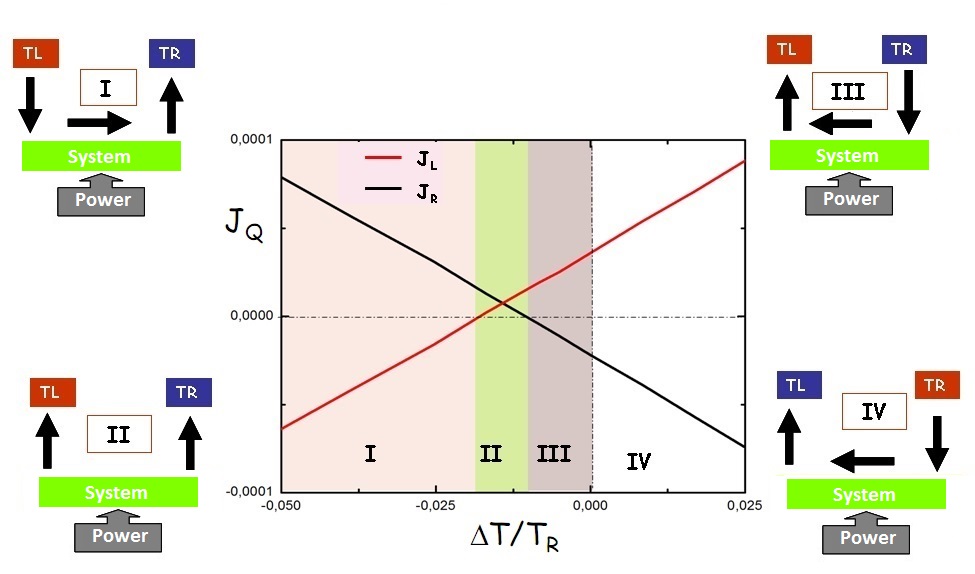} 
\caption{Central panel: Phase diagram $J_{L,R}$ versus $\Delta T/T_R$  for $\omega_0 =0.8$ ($\Delta T= T_R-T_L$).  Lateral panels sketch the flow direction in each regime: a) Regime I, b) Regime II, c) Regime III, d) Regime IV.   Parameters: $K_L=1=2 K_R, K_{int}=0.05$. }
\label{fig:Diagfase}
\end{figure}

From now on and without loss of generality, we take $T_R$ the reference temperature, being $T_L$ the variable one. We define $\Delta T =T_R - T_L$ and calculate $J_R$ and $J_L$ from Eq. \ref{FullJ}, where the sign for rectified currents is positive when heat flows from the system to the reservoir.   In Fig.\ref{fig:Diagfase} we show $J_{L,R}$ versus $\Delta T$ for a resonant frequency.

We find four heat transport regimes:

Regime $I$: $T_L > T_R$, $J_R>0$ and  $J_L < 0$
This is a regime mainly ruled by the temperature gradient, with heat flowing from the hot to the cold reservoir. 
The difference $J_R - J_L$ correspond to the mean power injected according to Eq. \ref{power}.
Although the currents have different absolute values, the power injected is low enough so the direction of the heat current is determined by $\Delta T$.  The system works as a heat conductor.
 
Regime $II$: $T_L > T_R$, $J_{R, L}>0$
This is a regime mainly governed by the injected power that is dissipated into both reservoirs.
The onset of this regime is given by a current reversal in one of the segments. When $J_R$ or $J_L$ are equal to zero the device can operate as a local insulator, thus heat is inhibited to flow in one part of the system.

Regime $III$: $T_L > T_R$, $J_R<0$ and  $J_L >0$
The system acts as a refrigerator pumping energy against temperature gradient. Heat flows from the cold reservoir to the hot one. 

Regime $IV$: $T_L < T_R$, $J_R<0$ and  $J_L >0$
Heat flows from hot to cold but in the reverse direction than Regime $I$. Transport is again mainly ruled by the temperature gradient. 

Depending on the driving frequency the system may display all or some of these regimes. Thus, adjusting temperature gradients and ac-frequencies it is possible to tune changeovers between them. 

\begin{figure}[h]
\hspace*{\fill}
\includegraphics[scale=0.17]{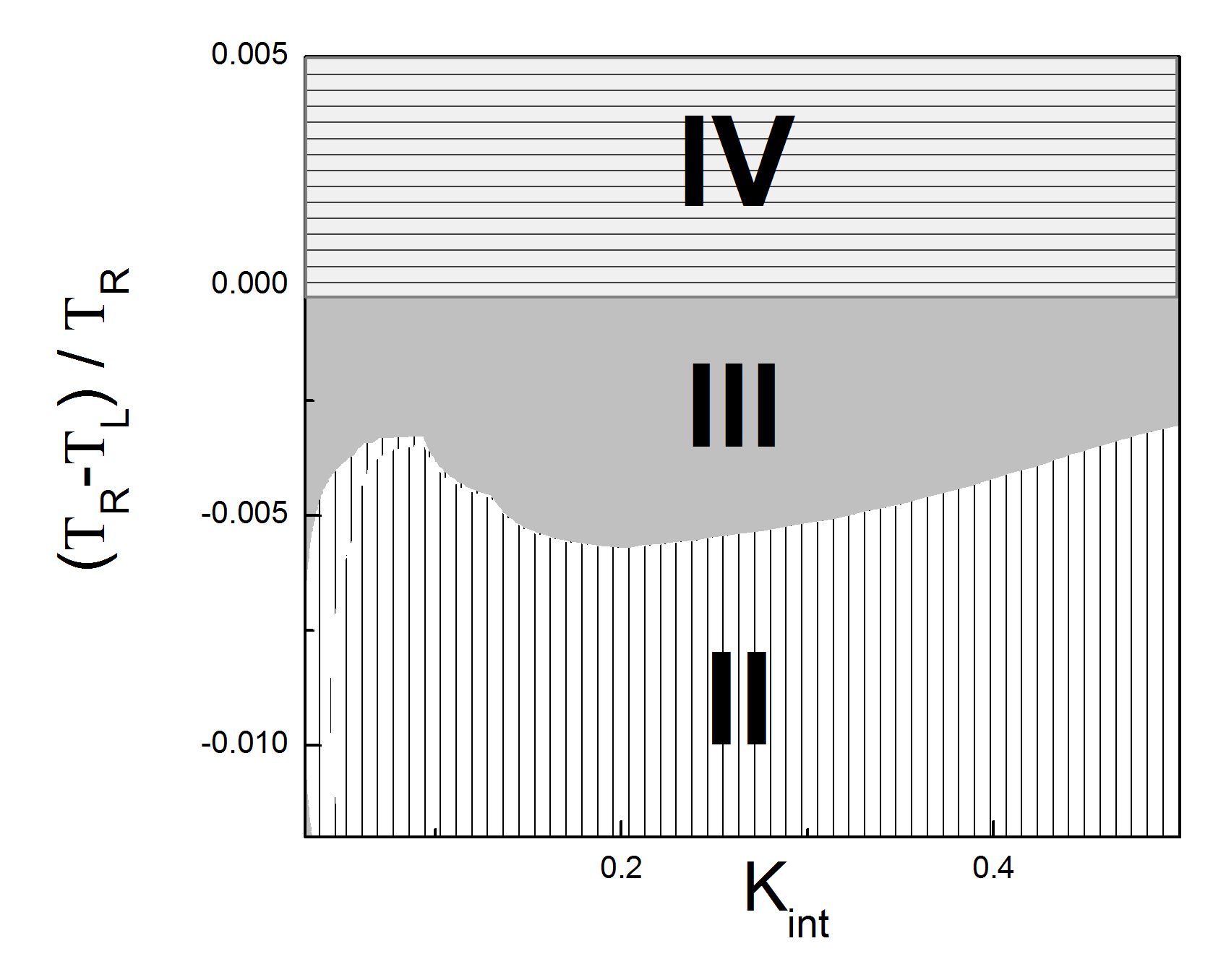} 
\hspace*{\fill}
\includegraphics[scale=0.17]{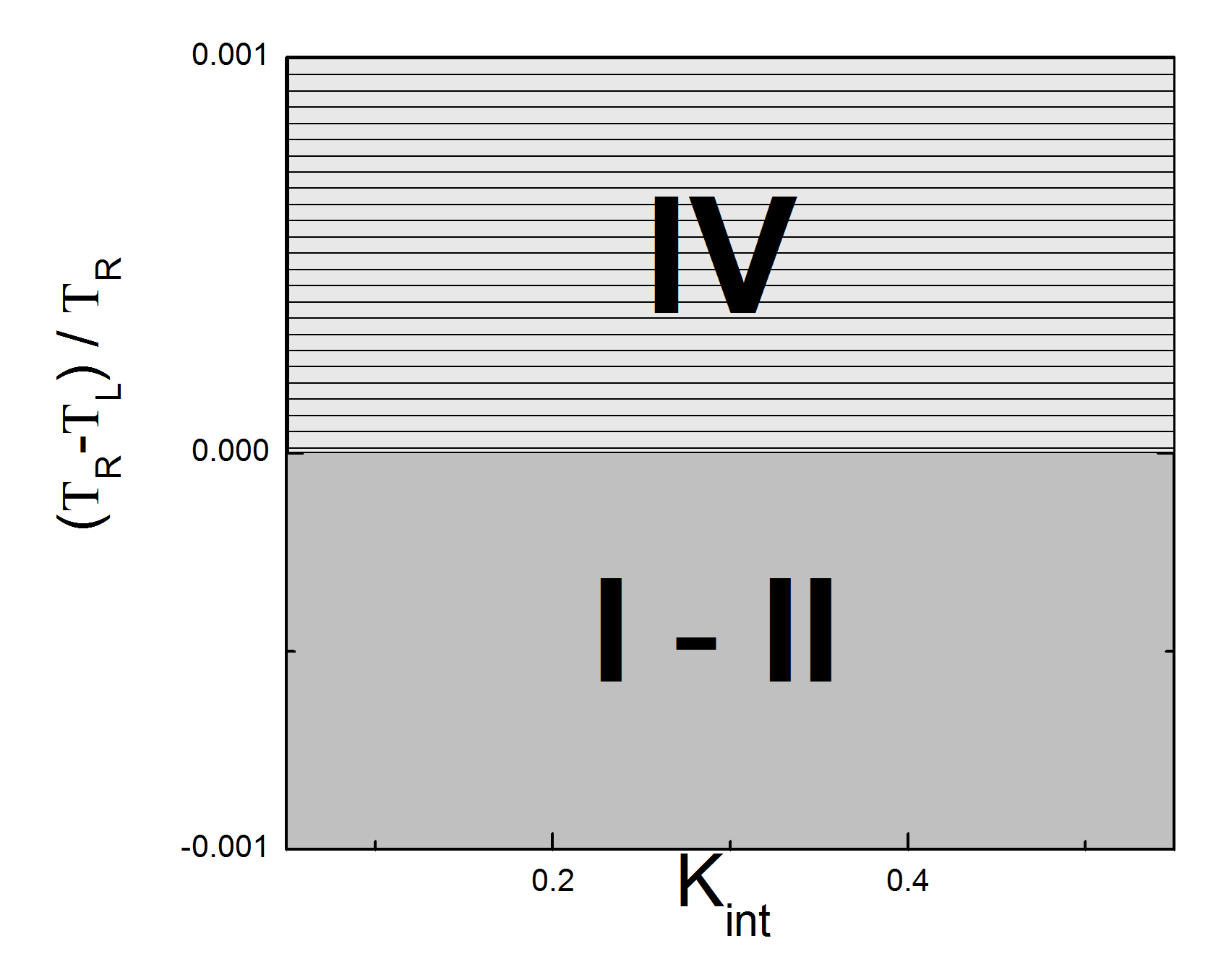}
\hspace*{\fill}
\caption{Left: $\omega_0 = 0.8$.  Right: $\omega_0 = 0.4$. Region $III$ corresponds to cooling phenomena. Parameters: same that in Fig.\ref{fig:Diagfase}.}
\label{fig:conturs}
\end{figure}

The transmission function of the system is strongly dependent on the structural parameters such as $K_{int}$,  who plays an important role on the potential regimes and transitions between them.  In Fig.\ref{fig:conturs} we show the contour phase diagram versus $\Delta T/T_R$ and $K_{int}$  for $\omega_0 = 0.8$ and $0.4$. In both cases the weak coupling condition is fulfilled and $T _k^{\pm}$ takes non zero values. Nevertheless we find that the system cannot pump energy against temperature bias in both cases.  For $\omega_0 = 0.8$ the refrigeration regime extends over a considerable region of the parameter space. However for $\omega_0 = 0.4$ this regime is absent.
As it was mentioned above, one possible reason is related to the differences between the transmission functions. In fig.\ref{fig:shift} we compare $T _{k=1}$ for both cases finding that high frequency channels associated to localized modes are not available for $\omega_0 = 0.8$. 

\begin{figure}[h]
\hspace*{\fill}
\includegraphics[scale=0.22]{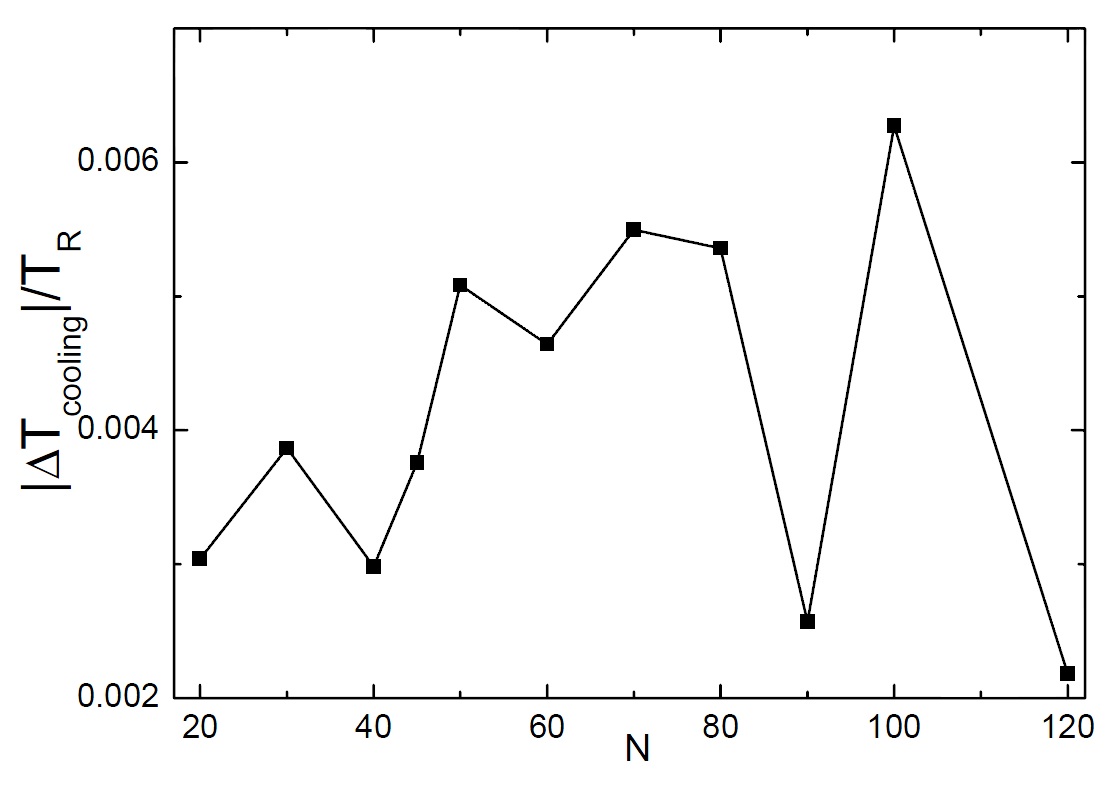} 
\hspace*{\fill}
\includegraphics[scale=0.22]{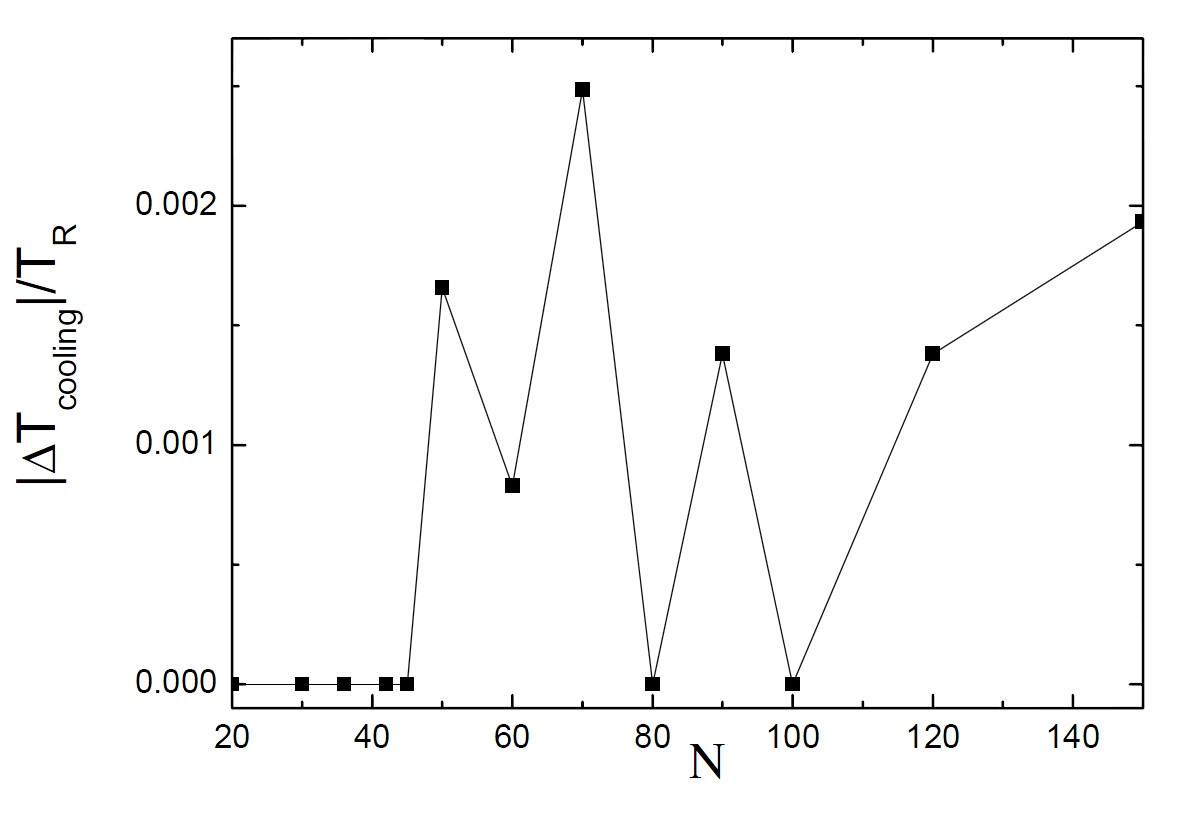} 
\hspace*{\fill}
\caption{$\Delta T_{cooling}$ versus $N$ for $\omega_0=0.8$ (left) and  $\omega_0=0.4$ (right). Zero values means that no cooling effect is possible.}
\label{fig:DeltaTRef}
\end{figure}

We also find another interesting feature that is related to the structure. For a reference value $T_R$, there is a temperature $T_{L_{min}}$ for which cooling regime is achived with a temperature difference between reservoirs is $\Delta T_{cooling}=T_R-T_{L_{min}}$. The last is strongly dependent on the size of the system, as it is shown in Fig.\ref{fig:DeltaTRef}. When $N$ is changed the spectral density also does, then the possible driving frequencies that will enable cooling in resonant condition in the parameter space.
From an experimental point of view it can be crucial to decide the way to increase the temperature intervals for which cooling phenomena occurs. One possibility is related to structural features or "phononic engineering", varying the size of the system as in our work or adjusting characteristic parameters associated to interactions between atoms (or in other words the adequate selection of materials). Other possibility is related to operational features of a device, for example tuning the strength or frequencies of the time-dependent perturbations or adjusting the differences of temperature along the device. Figs. \ref{fig:Diagfase}, \ref{fig:conturs} and \ref{fig:DeltaTRef} reveal the different possibilities.

\section{Performance of the phonon refrigerator}

The microscopic realization of a phonon refrigeration mechanism we are proposing is based on a {\it dynamical} cycle given by the periodic modulation of $K_{int}$ that produce the activation (or deactivation) of phonon modes or thermal channels.

Since the structure of the leads is asymmetric they produce different phonon speed in each segment. Additionally as they are finite, the system has only limited time before the reflected or transmitted phonons from the reservoirs begin to arrive and interfere with other phonons moving through the contact or propagate along it.  Therefore for certain driving frequencies and $\Delta T$ intervals, the system can block the usual heat flowing in the direction of the temperature gradient, enabling the pumping of phonons against it.

Unlike other works based on a time-dependent on-site potential \cite{AMCC}, our mechanism is based on an periodical contact interaction. Therefore our model can describe composite systems on substrates that can be realized by linear lattices with static pinning potentials. Examples of such systems are polymeric polar chains where the opening or locking of phonon channels can be adjusted under the action of electric fields and composed short alkane chains o molecular junctions that can be stretched or pulled under some defined protocol\cite{FET,stret1,stret2}.

As the underlying mechanism is the shift and changes in the width of the phonon bands, the system size $N$ is one of the structural parameters that plays a relevant role affecting the number of available phonon modes, and therefore the cooling ability of the device.  We calculate the ratio of cooling to energy consumption or Cooling Coefficient of Performance $CP$ as:

\begin{equation}
CP = \frac{Q_{cold}}{Q_{hot}-Q_{cold}}
\label{eq:cop}
\end{equation}

where $Q_{cold}$ is the heat extracted from the cold reservoir and $Q_{hot}$ is the heat injected on the hot one. The energy consumption or power injected is thus given by $Q_{hot}-Q_{cold}$.

\begin{figure}
\centering
\includegraphics[scale=0.3]{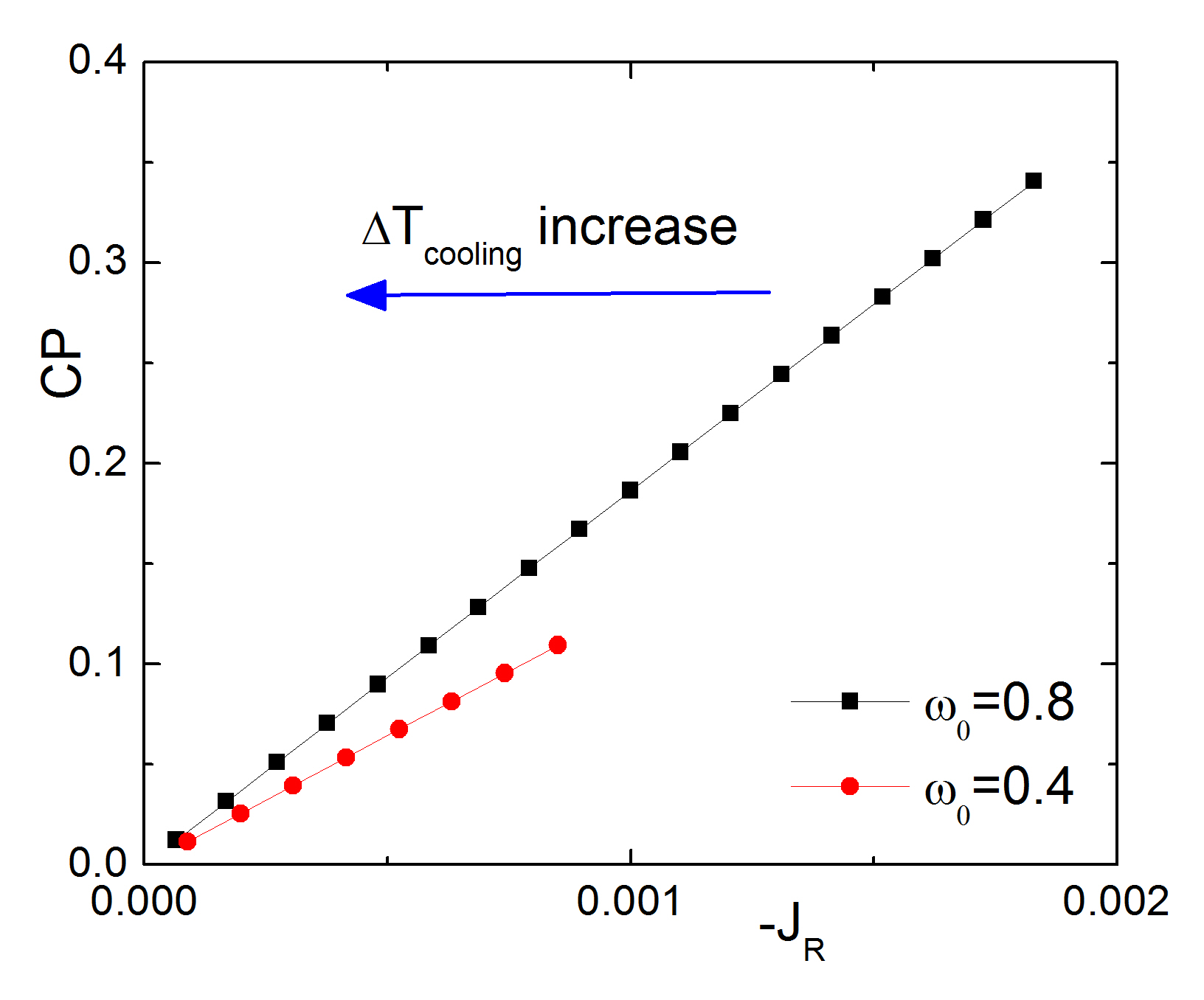}
\caption{Plot of the cooling efficiency versus the heat pumped from the colder reservoir for driving frequencies $\omega_0=0.8$ and $0.4$ with $N=150$.}
\label{fig:CompEf}
\end{figure}

In Fig.\ref{fig:CompEf} we show the performance of the device plotting $CP$ as a function of the heat pumped from the cooler reservoir. In our system $\omega_0$ is an operational control parameter for the efficiency when the device is subject to a temperature difference. The curves show the expected behaviour: as long the heat extracted is bigger so the efficiency of the device. However, the efficiency presents an upper and lower bound for the heat that can be extracted. The maximun and minimun heat extracted for an $\omega_0$ correspond to a $\Delta T_{cooling}$ going from zero to the maximun value respectively obtained from the phase diagram. 
In addition, the $CP$ presents a monotonic behaviour, decreasing faster as long the heat is lower.  

\begin{figure}
\centering
\includegraphics[scale=0.3]{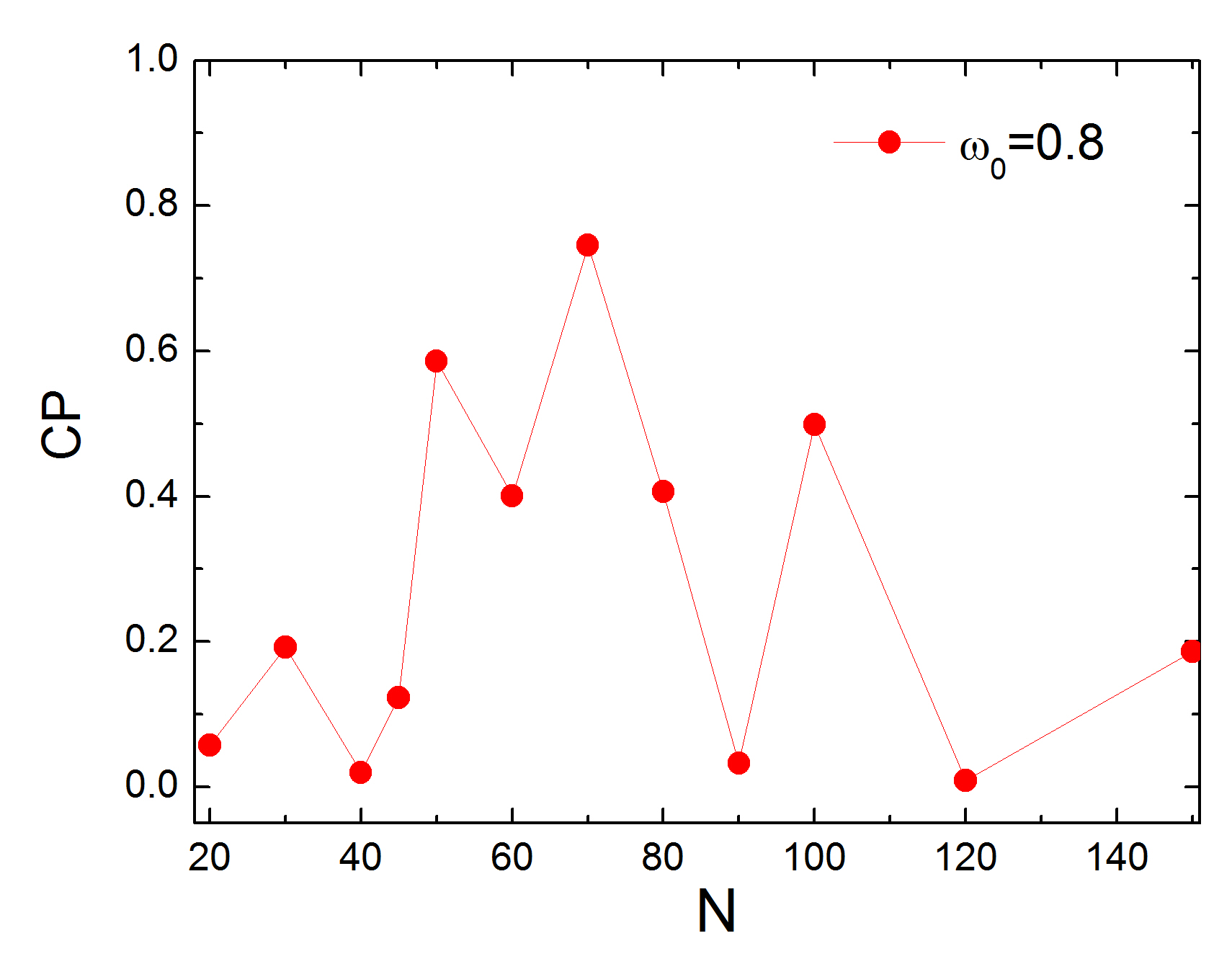}  
\caption{$CP$ vs $N$ for $\omega_0=0.8$.}
\label{fig:COP}
\end{figure}

The multiresonant feature of the transport is also reflected in the cooling performance of the system as it is shown in Fig.\ref{fig:COP} where $CP$ is plotted versus the size of the system. Therefor a similar non monotonic response can be also obtained by tuning adequately other structural features related to material properties (interaction constants) or operational factors.

\section{Conclusions}

We analyzed the heat current in a microscopic system consisting in a one-dimensional finite composed chain of atoms. It is built from two segments interacting with a substrate. They interact periodically by one of their ends, meanwhile the others are connected to phononic reservoirs. Therefore, our model considered heat transfer assisted by 'longitudinal' vibrational modes.  We have calculated the current using a nonequilibrium Green's function approach, treatment capable of handling time-dependent perturbations and temperature gradients at the same time. 

We found that the system displays different heat transfer regimes according to the temperature gradients, structural  and operational factors.
In one of them the device can act as a refrigerator, pumping energy against temperature gradient assisted by the power injected into the system.  The underlying mechanism is based in the periodic modulation of the contact between different structures of the composite that opens or closes local phonon channels, enabling or suppressing the propagation of phonons along the interface.
We showed two ways to increase or reduce the performance a system to operate as a heat pump under a temperature difference. One possibility is to design a device from a given material with the adequate size when the external action is fixed.  Other possibility is to exert a controlled external action in order to tune the available thermal channels.

Nowadays, the energies involved in the micro and nanolevel applications as molecular electronics or nanomechanical devices can be important even producing damages or reduction of the operation performance of the object. We explore and proposed a physical mechanism for temperature reduction and control of the local heat flow along a device, topic of increasing and actual technological relevance.

\section*{Acknowledgments}
M.F.C., N.B., A.S are supported by PIO Conicet. 
We thank to cluster TUPAC of the Computational Simulation Center - CONICET.

\end{document}